\documentclass[conference]{IEEEtran}
\IEEEoverridecommandlockouts
\usepackage{cite}
\usepackage{amsmath,amssymb,amsfonts,latexsym}
\usepackage{algorithmic}
\usepackage{graphicx}
\usepackage{subcaption}
\usepackage{makecell}
\usepackage{textcomp}
\usepackage{xcolor}
\def\BibTeX{{\rm B\kern-.05em{\sc i\kern-.025em b}\kern-.08em
    T\kern-.1667em\lower.7ex\hbox{E}\kern-.125emX}}

\usepackage{mathtools}
\usepackage{comment}
\usepackage{hyperref}
\usepackage{booktabs} 
\usepackage{url}            

\ifCLASSINFOpdf

\else

\fi

\usepackage{amssymb}
\usepackage{amsmath}
\usepackage{titlesec}
\usepackage{url}
\usepackage{xcolor}

\begin{document}
%
\title{It's-A-Me, Quantum Mario: Scalable Quantum Reinforcement Learning with Multi-Chip Ensembles 
\thanks{The views expressed in this article are those of the authors and do not represent the views of Wells Fargo. This article is for informational purposes only. Nothing contained in this article should be construed as investment advice. Wells Fargo makes no express or implied warranties and expressly disclaims all legal, tax, and accounting implications related to this article.}
}

\author{
Junghoon Justin Park$^1$, Huan-Hsin Tseng$^2$, Shinjae Yoo$^2$, Samuel Yen-Chi Chen$^3$, Jiook Cha$^{1,4,5}$ \\
\small $^1$Interdisciplinary Program in Artificial Intelligence, Seoul National University $\quad$ \\ $^2$ Computational Science Initiative, Brookhaven National Laboratory $\quad$ \\ $^3$ Wells Fargo $\quad$ \\$^4$ Department of Psychology, Seoul National University $\quad$ \\ $^5$ Department of Brain and Cognitive Sciences, Seoul National University $\quad$ \\
\small \texttt{utopie9090@snu.ac.kr, \{htseng, sjyoo\}@bnl.gov, yen-chi.chen@wellsfargo.com, connectome@snu.ac.kr} 
}



\maketitle

\begin{abstract}
Quantum reinforcement learning (QRL) promises compact function approximators with access to vast Hilbert spaces, but its practical progress is slowed by NISQ-era constraints such as limited qubits and noise accumulation. We introduce a multi-chip ensemble framework using multiple small Quantum Convolutional Neural Networks (QCNNs) to overcome these constraints. Our approach partitions complex, high-dimensional observations from the Super Mario Bros environment across independent quantum circuits, then classically aggregates their outputs within a Double Deep Q-Network (DDQN) framework. This modular architecture enables QRL in complex environments previously inaccessible to quantum agents, achieving superior performance and learning stability compared to classical baselines and single-chip quantum models. The multi-chip ensemble demonstrates enhanced scalability by reducing information loss from dimensionality reduction while remaining implementable on near-term quantum hardware, providing a practical pathway for applying QRL to real-world problems.
\end{abstract}


%
\IEEEpeerreviewmaketitle

\section{\textbf{Introduction}}
Quantum reinforcement learning (QRL) has emerged as a promising frontier integrating quantum computing capabilities with reinforcement learning paradigms \cite{Chen, Jerbi, Skolik, Park2024}. Recent advances have demonstrated viable QRL implementations through variational quantum circuits (VQC), with Chen et al. (2020) pioneering a quantum analog to deep Q-learning networks by substituting neural networks with VQCs \cite{Chen}. Subsequent research has refined these approaches, validating that quantum models can effectively learn control policies through value iteration on standard benchmarks \cite{Skolik, Lockwood2020}. These developments highlight QRL's potential to leverage quantum parallelism and entanglement for enhanced learning efficiency \cite{Park2024}.

Nevertheless, QRL still faces severe scalability challenges. The limitations of noisy intermediate-scale quantum (NISQ) devices \cite{Bharti} have confined most QRL demonstrations to simplified problems using shallow circuits with only 4-6 qubits \cite{Chen, Jerbi, Skolik}. While it is possible to process high-dimensional data with limited number of qubits using amplitude encoding, it requires exponential circuit depth \cite{Plesch, Sun2023, Zhang2021}. As quantum noise grows exponentially with circuit depth \cite{DePalma}, large-scale quantum circuits are often infeasible. Consequently, current quantum agents operate with restricted state representations and can only tackle low-dimensional environments. Even recent advancements remain limited to elementary tasks such as CartPole \cite{Lockwood2020, Lockwood2021}. The fundamental challenge lies in scaling to high-dimensional spaces, which demands larger quantum circuits that inevitably encounter deeper noise effects and the barren plateau problem—the vanishing of gradients in large quantum circuits \cite{McClean, Larocca}. This scalability bottleneck prevents QRL from addressing complex environments where quantum advantages might be most significant.

Distributed quantum computing techniques offer a promising avenue to overcome these limitations. While extensively explored in quantum computing broadly \cite{Kimble, Peng, Barral}—including feature-based partitioning approaches that distribute input data across multiple circuits with classical aggregation \cite{Kawase}—these strategies have remained largely unexplored in QRL \cite{Wu, Pira, QUDIO}. This gap represents a missed opportunity to scale quantum agents to more challenging domains.

In this work, we introduce a scalable QRL framework leveraging multi-chip ensembles. The multi-chip ensemble architecture has been shown to offer advantages including greater scalability, barren plateau mitigation, improved generalization, and noise resilience \cite{Park2025}. We apply this modular approach by partitioning the agent's quantum policy model across multiple quantum processing units and classically aggregating their outputs for decision-making. Specifically, we divide observation data into subsets, process them through parallel smaller quantum circuits on separate quantum processors, then combine their expectation values to produce final Q-value estimates or policy outputs. Drawing inspiration from classical ensemble learning and recent distributed VQC method, our approach effectively constructs a larger virtual quantum model from NISQ-compatible components. We demonstrate our approach's efficacy by training a quantum agent to play Super Mario Bros \cite{SuperMarioBros}—a complex, high-dimensional environment previously beyond the reach of traditional QRL methods.

\section{\textbf{Background}}
\subsection{Reinforcement Learning}\label{Backgroun_RL}
RL is a framework where an agent learns to maximize cumulative rewards through interactions with an environment. This process is formalized as a Markov Decision Process (MDP) defined by the tuple $\langle \mathcal{S, A, R, P,}\gamma \rangle$ where $\mathcal{S}$ represents the state space, $\mathcal{A}$ the action space, $\mathcal{R}$ the reward function, $\mathcal{P}$ the transition probability function, and $\gamma \in [0,1)$ the discount factor.

The agent's objective is to find an optimal policy $\pi^*$ that maximizes expected future rewards. This is quantified by the action-value function $Q^{\pi}(s,a)$, which represents the expected return when taking action $a$ in state $s$ and following policy $\pi$ thereafter:
\begin{align}
Q^{\pi}(s,a) = \mathbb{E} \Bigg[ \sum_{t=0}^{\infty} \gamma^t R(s_t,a_t) \Bigg], \\
s_t \sim P(s_{t-1},a_{t-1}), \quad a_t \sim \pi(s_t) \nonumber
\end{align}
This function satisfies the Bellman equation~\cite{Bellman}:
\begin{equation}
Q^{\pi}(s,a) = \mathbb{E}\Big[\mathcal{R}(s,a)\Big] + \gamma \, \mathbb{E}_{\mathcal{P}, \pi} \Big[Q^{\pi}(s', a')\Big] 
\end{equation}

Q-learning, a fundamental RL algorithm, iteratively updates the action-value function using temporal difference learning:
\begin{multline}
    Q(s_t,a_t) \leftarrow Q(s_t,a_t) + \alpha \left( r_{t+1} + \gamma \max_a Q(s_{t+1},a) \right. \\ \Big. - Q(s_t, a_t) \Big)
\end{multline}
where $\alpha$ is the learning rate and $r_{t+1}$ the immediate reward.

Deep Q-Networks (DQN)~\cite{Mnih} extend Q-learning by approximating the action-value function with neural networks. Double DQN (DDQN)~\cite{vanHasselt, vanHasseltHado} improves upon standard DQN by addressing its overestimation bias through decoupling action selection and evaluation. DDQN employs two networks: an online network with parameters $\theta$ for action selection and a target network with parameters $\bar{\theta}$ for value estimation. The loss function is:
\begin{multline}
    L(\theta) = \mathbb{E}_{(s,a,r,s') \sim \mathcal{D}} \Big[ \big( r + \gamma Q_{\bar{\theta}}(s', \arg\max_{a'} Q_{\theta}(s',a')) \\
    - Q_{\theta}(s,a) \big)^2 \Big]
\end{multline}
where $\mathcal{D}$ is a replay buffer of transitions $(s,a,r,s')$. By separating action selection (via $\arg\max_{a'} Q_{\theta}$) from value estimation (via $Q_{\bar{\theta}}$), DDQN significantly reduces the systematic overestimation present in standard Q-learning with function approximation.

\subsection{VQCs in RL}
VQCs offer a quantum approach to function approximation in RL, analogous to neural networks in classical RL~\cite{Benedetti}. A VQC consists of three key components: an encoding layer that transforms classical states into quantum states through rotational gates, a parameterized variational ansatz that processes quantum information, and a measurement layer that extracts expectation values to represent action-values.

In QRL, VQCs replace neural networks in DQN frameworks~\cite{Chen, Jerbi, Lockwood2020, Skolik, Park2024}. The process operates as follows: first, the environment state $s$ is encoded into a quantum state through rotation gates applied to initialized qubits. Next, the encoded state passes through a parameterized quantum circuit (the ansatz) whose parameters $\boldsymbol{\theta}$ are trainable. Finally, Pauli operators measure each qubit (or combinations thereof) to produce expectation values that correspond to Q-values for each possible action:
\begin{equation}
Q_\theta(s,a) = \text{Tr}[H_a\rho(s,\theta)]
\end{equation}
where $\rho(s,\theta) = |\psi(s,\theta)\rangle\langle\psi(s,\theta)|$ is the density matrix representing the quantum state after encoding and applying the variational circuit, and $H_a$ is the Hermitian operator corresponding to action $a$. This formulation generalizes to both pure and mixed quantum states, making it particularly relevant for NISQ devices.

The circuit parameters $\boldsymbol{\theta}$ are updated by minimizing a loss function equivalent to that used in classical DQN or DDQN, with gradients computed through parameter-shift rules \cite{Schuld2019} or other quantum-compatible optimization methods \cite{Chen, Skolik}.

This VQC-based approach to DQN potentially offers advantages through quantum phenomena such as superposition and entanglement, while remaining compatible with existing RL algorithms and frameworks \cite{Park2024}.

\subsection{Multi-chip Ensembles}
Multi-chip ensembles represent a distributed quantum computing approach that addresses scalability limitations in VQCs~\cite{Park2025}. This architecture combines $k$ separate $l$-qubit quantum processors to effectively simulate a larger $n$-qubit system ($n = k \times l$) without requiring physical qubit connectivity between chips.

The architecture is characterized by a tensor product of independent subcircuits:
\begin{equation}
U_{\text{MC}}(\boldsymbol{\theta}) = \bigotimes^k_{i=1} U_i(\boldsymbol{\theta}_i)
\end{equation}
where each $U_i(\boldsymbol{\theta}_i)$ operates on a separate quantum chip with its own parameter set $\boldsymbol{\theta}_i$. 

For data processing, an input vector $\boldsymbol{x} \in \mathbb{R}^n$ is partitioned into $k$ subvectors $\boldsymbol{x} = [\boldsymbol{x}_1, \boldsymbol{x}_2, \dots, \boldsymbol{x}_k]$, where each $\boldsymbol{x}_i \in \mathbb{R}^\ell$ is processed by the corresponding quantum chip. Each subcircuit independently encodes its subvector, applies parameterized operations, and produces an expectation value:
\begin{equation}
f_{\boldsymbol{\theta}_i}(\boldsymbol{x}_i) = \text{Tr}[H_i U_i(\boldsymbol{\theta}_i) \rho_i(\boldsymbol{x}_i) U_i^\dagger(\boldsymbol{\theta}_i)]
\end{equation}
where $\rho_i(\boldsymbol{x}_i) = V_i(\boldsymbol{x}_i) \left( {}^{\otimes \ell}|0\rangle \langle0|^{\otimes \ell} \right) V^\dagger_i(\boldsymbol{x}_i)$ denotes the encoded quantum state with $V_i(\boldsymbol{x}_i)$ representing the data encoding unitary for the $i$-th subcircuit, and $H_i$ is the measured observable.

These individual measurements are combined through a classical function $g: \mathbb{R}^k \to \mathbb{R}^m$:
\begin{equation}
f_{\boldsymbol{\theta}}(\boldsymbol{x}) = g(f_{\boldsymbol{\theta}_1}(\boldsymbol{x}_1), \ldots, f_{\boldsymbol{\theta}_k}(\boldsymbol{x}_k)).
\end{equation}
Finally, the parameters $\boldsymbol{\theta} = \{\boldsymbol{\theta}_1, ..., \boldsymbol{\theta}_k\}$ are optimized jointly to minimize a task-specific loss function:
\begin{equation}
\mathcal{L}(\boldsymbol{x}, y; \boldsymbol{\theta}) = \mathcal{L}_{\text{ensemble}}(f_{\boldsymbol{\theta}}(\boldsymbol{x}), y).
\end{equation}

Notably, this architecture sacrifices cross-chip entanglement in exchange for three critical advantages \cite{Park2025}: (1) it enables processing higher-dimensional data than would be possible on with limited number of physical qubits; (2) it mitigates barren plateau problems by using shallower subcircuits; and (3) it allows parallel computation of gradients during training, improving efficiency as the system scales.

\section{\textbf{Scalable QRL with Multi-chip Ensembles}}
We introduce a scalable QRL framework that leverages multi-chip ensemble architecture to overcome the dimensionality limitations of current QRL methods. Our approach, which we demonstrate on the complex Super Mario Bros environment \cite{SuperMarioBros}, combines DDQN with distributed quantum processing to handle high-dimensional state spaces efficiently.

\begin{figure}[htbp]
    \centering
    \includegraphics[width = 1\columnwidth]{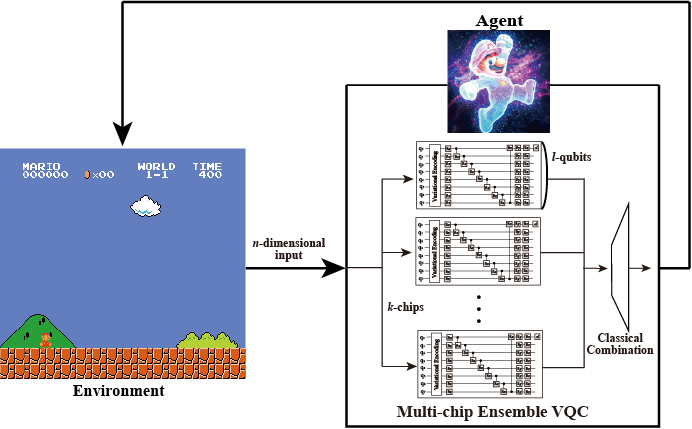}
    \caption{\textbf{Multi-chip Ensemble Architecture for QRL}. The $n$-dimensional input state is partitioned into $k$ separate feature vectors. Each vector is processed independently by an $l$-qubit VQC on a separate simulated quantum chip (i.e., $n = k \times l$). The VQCs apply variational quantum operations, maintaining entanglement only within each individual chip (no inter-chip quantum connections). Measurement outcomes from all chips are classically aggregated through a learnable function to produce action Q-values for the DDQN algorithm. This distributed architecture enables processing of higher-dimensional feature representations than would be possible with a single quantum circuit, reducing information loss from dimensionality reduction while remaining implementable on NISQ-era hardware.}
    \label{fig_MultiChip_QRL}
\end{figure}

\subsection{Quantum Convolutional Neural Networks for RL}
Quantum Convolutional Neural Networks (QCNNs) extend classical CNNs to the quantum domain, offering advantages in parameter efficiency and noise resilience~\cite{Cong}. QCNNs follow a hierarchical structure of alternating convolutional and pooling layers of quantum operations. The convolutional layer applies local parameterized unitaries to overlapping regions of qubits. The pooling layer reduces the quantum state dimension through partial measurements.

A critical advantage of QCNNs is their logarithmic parameter scaling—requiring only $O(log(n))$ variational parameters for an $n$-qubit system~\cite{Cong}. Importantly, QCNNs exhibit provable resilience against barren plateaus, maintaining non-vanishing gradients even as system size increases~\cite{Pesah}. This property ensures trainability on near-term quantum devices where gradient-based optimization is essential.

Additionally, QCNNs demonstrate potential quantum advantages in specific image recognition tasks through efficient encoding and processing of structured data~\cite{Kerenidis}. These characteristics—parameter efficiency, barren plateau resistance, and structured data processing—make QCNNs an ideal candidate for our multi-chip ensemble approach to QRL.

\subsection{Conventional Quantum Approach: Single-chip QCNN Architecture}
We implement quantum DDQN by replacing the classical deep neural network with a QCNN. This section describes the single-chip implementation, which serves as the foundation for our multi-chip approach.

The QCNN processes environment states through a hierarchical structure of quantum operations. For a given state $\boldsymbol{s} \in \mathcal{S}$, we first encode it into an $n$-qubit quantum state represented by density matrix $\boldsymbol{\rho}_s$.

The QCNN architecture consists of $L$ sequential layers, each comprising a convolutional and a pooling operation. The convolutional layer applies parameterized unitary transformations across qubit groups:
\begin{equation}
   \mathcal{U}^{(l)}(\boldsymbol{\theta})(\boldsymbol{\rho}) = \Big(\bigotimes^{m_l}_{j=1}U_j^{(l)}(\theta_j^{(l)})\Big)\boldsymbol{\rho}\Big(\bigotimes^{m_l}_{j=1}U_j^{(l)}(\theta_j^{(l)})\Big)^{\dagger}
\end{equation}
where $m_l$ is the number of local unitaries in layer $l$, and each $U_j^{(l)}(\theta_j^{(l)})$ is a parameterized quantum gate acting on a subset of qubits. The pooling layer reduces the quantum system dimension through partial trace operations:
\begin{equation}
\mathcal{P}^{(l)}(\boldsymbol{\rho}) = \text{Tr}_\text{discard}[\boldsymbol{\rho}]
\end{equation}
where ``discard'' indicates qubits being traced out.

These operations combine to form a quantum channel for each layer:
\begin{equation}
\mathcal{C}^{(l)}(\boldsymbol{\rho}) = \mathcal{P}^{(l)}\big(\mathcal{U}^{(l)}(\boldsymbol{\theta})(\boldsymbol{\rho})\big).
\end{equation}
Starting with $\boldsymbol{\rho}^{(0)}=\boldsymbol{\rho}_s$, each layer transforms the state:
\begin{equation}
\boldsymbol{\rho}^{(l)}=\mathcal{C}^{(l)}\big(\boldsymbol{\rho}^{(l-1)}\big) \quad \text{for} \quad l=1,2,...,L
\end{equation}
After processing through all $L$ layers, we obtain the final state:
\begin{equation}
\boldsymbol{\rho}^{(L)}= \Big(\prod^L_{l=1} \mathcal{C}^{(l)}\Big)(\boldsymbol{\rho}^{(0)})
\end{equation}

To compute Q-values for DDQN, we measure action-specific observables $H_a$ on the final state:
\begin{equation}
Q_{\boldsymbol{\theta}}(s,a) = \text{Tr}\big[H_a\boldsymbol{\rho}^{(L)}\big]
\end{equation}

This quantum-derived Q-function is then used in the standard DDQN algorithm, following the same training procedure detailed in Section \ref{Backgroun_RL}, but with quantum circuit parameters $\boldsymbol{\theta}$ as trainable weights rather than classical neural network parameters.

\subsection{Multi-chip Ensemble QCNN Architecture}
We extend our QRL framework by distributing computation across multiple quantum processors through a multi-chip ensemble approach (Fig. \ref{fig_MultiChip_QRL}). This architecture addresses the fundamental scalability limitations of single-chip quantum circuits when processing high-dimensional game environments.

For an input state $\boldsymbol{s} \in \mathcal{S}$, we partition the data into $k$ subvectors:
\begin{equation}
\boldsymbol{s} = [\boldsymbol{s}_1, \boldsymbol{s}_2, \ldots, \boldsymbol{s}_k]
\end{equation}
where each $\boldsymbol{s}_i$ is processed by a separate quantum chip.

Each quantum chip $i$ independently:
\begin{enumerate}
    \item Encodes its subvector into an initial quantum state $ \boldsymbol{\rho}^{(0)}_i$
    \item Processes this state through $L$ layers of convolutional and pooling operations:
    \begin{equation}
    \boldsymbol{\rho}^{(L)}_i = \Big(\prod^L_{l=1} \mathcal{C}^{(l)}_i\Big)(\boldsymbol{\rho}^{(0)}_i)
    \end{equation}
    \item Measures an action-specific observable to produce a scalar output:
    \begin{equation}
    f_{\boldsymbol{\theta}_i}(\boldsymbol{s}_i) = \text{Tr}[H_{a_i}\boldsymbol{\rho}^{(L)}_i]
    \end{equation}
\end{enumerate}

This architecture offers three key advantages over conventional single-chip QRL models:

First, it addresses the dimensionality barrier. While standard variational encoding requires $n$-qubits to process $n$-dimensional data (often necessitating lossy dimension reduction techniques), our approach distributes the computational load across $k$ independent quantum processors, each handling only $n/k$ dimensions of the input data.

Second, it avoids the limitations of amplitude encoding, which theoretically enables encoding $2^n$-dimensional data with $n$-qubits but requires deep circuits and complex state preparation infeasible for NISQ hardware~\cite{Plesch, Sun2023, Zhang2021}.

Third, it enables horizontal scaling through additional chips rather than requiring larger single quantum processors. For example, an ensemble of $200$ small $10$-qubit circuits can effectively process $2000$-dimensional data, far beyond what any near-term single quantum processor could handle. This approach makes high-dimensional RL environments like Super Mario Bros tractable for quantum processing without mandatory dimensionality reduction.

\section{\textbf{Experiment}}
We evaluate our multi-chip ensemble approach to quantum reinforcement learning using the Super Mario Bros environment from OpenAI Gymnasium~\cite{SuperMarioBros}, comparing three agent configurations: a classical baseline, a single-chip QCNN, and our proposed multi-chip ensemble QCNN.

\subsection{Environment and Preprocessing}
The agent's objective is to navigate Mario through game levels to reach the flag while maximizing score. The environment provides raw game screen images as observations. We preprocess these images to grayscale, resize to 84×84 pixels, and stack four consecutive frames to capture temporal information, resulting in a $28,224$-dimensional state vector (4×84×84). We simplify the action space to two discrete actions: ``walk right" and ``jump right" to focus on fundamental navigation skills.

\begin{figure*}[htbp]
    \centering
    \includegraphics[width = 2.0\columnwidth]{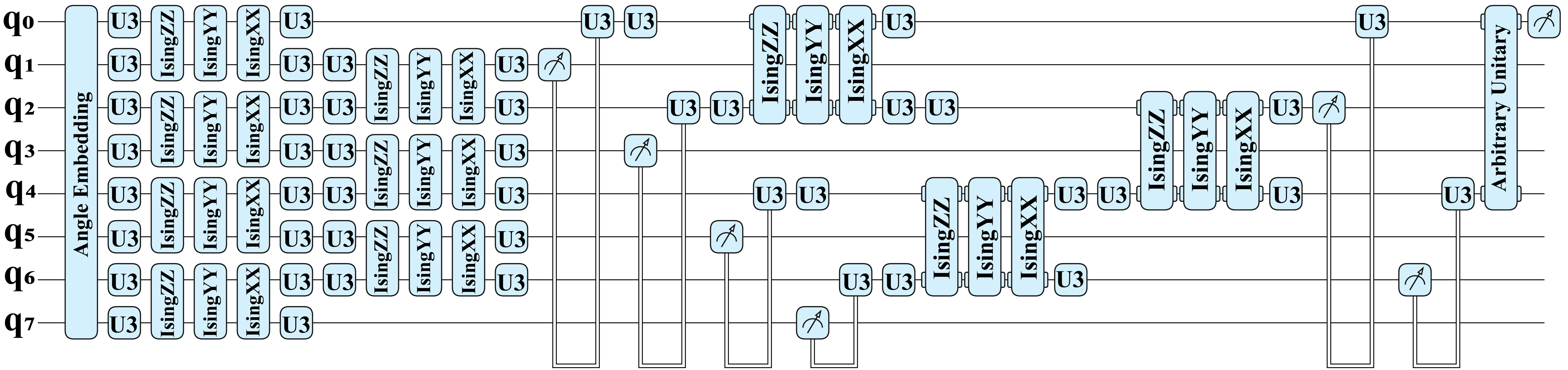}
    \caption{\textbf{Quantum Circuit Architecture for Individual QCNN Chip}. Each 8-qubit QCNN processor (used in both single-chip and multi-chip configurations) implements a two-layer hierarchical structure. Classical features are encoded via angle embedding, followed by convolutional layers comprising trainable U3 rotations on each qubit and IsingXX/YY/ZZ entangling gates between adjacent pairs. Pooling operations trace out every second qubit, reducing the system from 8 qubits to 4 qubits (Layer 1) to 2 qubits (Layer 2). A final unitary operation precedes measurement to extract expectation values for Q-value computation.}
    \label{fig_QCNN}
\end{figure*}

\subsection{Model Architectures}
All three models share a common DDQN framework but differ in their feature processing components. The classical baseline implements a conventional CNN-based architecture with fully connected layers that mirror the hierarchical processing of quantum circuits, providing a fair comparison benchmark.

The single-chip QCNN utilizes a hybrid architecture where initial classical linear layers perform substantial dimensionality reduction from $28,224$ to $n$ features, where $n$ is the number of qubits available on a single quantum processor. These features are then processed through a QCNN circuit, with final classical layers transforming the quantum measurements into Q-values (Figure \ref{fig_QCNN}).

Our proposed multi-chip ensemble QCNN employs a distributed approach wherein initial classical linear layers reduce dimensionality to $k \times l$ features, where $k$ is the number of quantum chips and $l$ is the number of qubits per chip. These features are partitioned across $k$ separate quantum circuits, each processing $l$ features. The final classical layers combine measurements from all quantum circuits and produce Q-values from the aggregated measurements. The key advantage of this architecture is its ability to process larger intermediate representations ($k \times l \gg n$) by distributing computation, thereby reducing information loss during the initial dimensionality reduction.

\subsection{Implementation Details}
We implemented all three models using PennyLane~\cite{bergholm2022pennylane} for quantum simulations integrated with PyTorch for classical processing. To ensure fair comparison, we maintained consistent hyperparameters across all agents: discount factor $\gamma=0.9$, learning rate $\alpha=0.00025$ with Adam optimizer, batch size of 32, and an exploration rate decay of 0.99999999 for the $\epsilon$-greedy policy.

For the quantum architectures, the single-chip QCNN used 8 qubits on a single simulated quantum processor. Our multi-chip ensemble QCNN distributed computation across $k$ independent subcircuits, where $k$ was varied (2, 10, 50, or 100) to investigate the impact of ensemble size. Each subcircuit comprised 8 qubits and an identical structure. This ensemble approach enabled the processing of an $8k$-dimensional intermediate representation, in contrast to the 8-dimensional representation in the single-chip model. As an aggregation method, the scalar expectation value from each of the $k$ chips was concatenated into a vector of size $k$, which was then processed by a trainable classical feed-forward network to produce the final Q-values.

Both single-chip and multi-chip ensemble QCNN models employed a two-layer architecture, with each layer consisting of a convolutional operation followed by a pooling operation. The classical baseline maintained an equivalent network structure. Specifically, its architecture replaces each 8-qubit QCNN with a classical module of fully connected layers designed to mimic the QCNN's hierarchical, two-layer processing flow, ensuring a structurally analogous comparison.

All experiments were conducted on a Linux server with 128 CPU cores (256 threads), 503 GB RAM, and an NVIDIA A100-PCIE GPU with 40 GB memory, using Python 3.11.7, PyTorch 2.5.0+cu121, and CUDA 12.1.

\section{\textbf{Results}}

\begin{figure*}[!ht]
    \centering
    \begin{subfigure}[ht]{0.5\textwidth}
        \centering
        \includegraphics[height=2.05in]{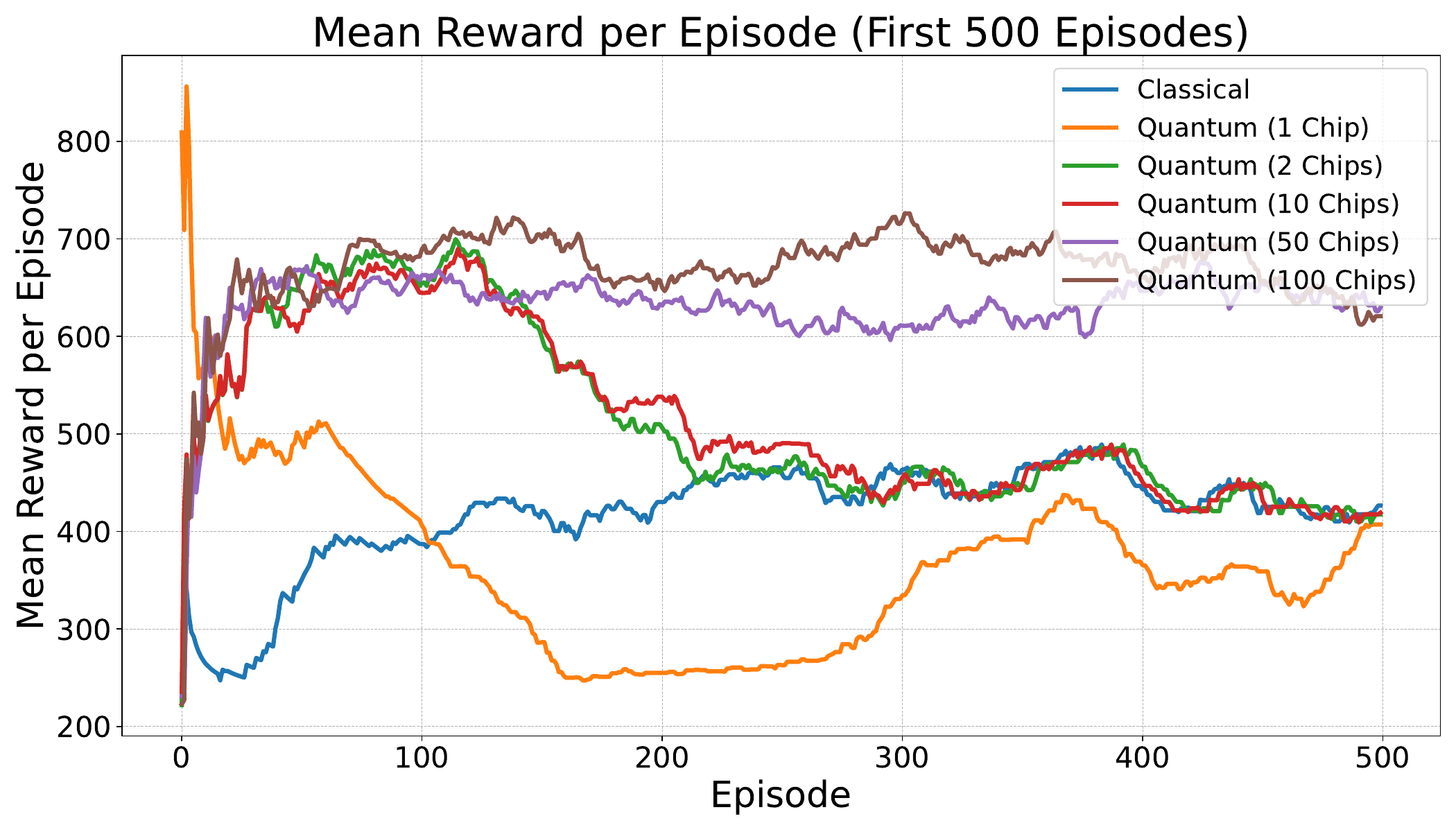}
        \caption{Mean Rewards}
        \label{Figure_Rewards}
    \end{subfigure}%
    \hfill    
    \begin{subfigure}[ht]{0.5\textwidth}
        \centering
        \includegraphics[height=2.05in]{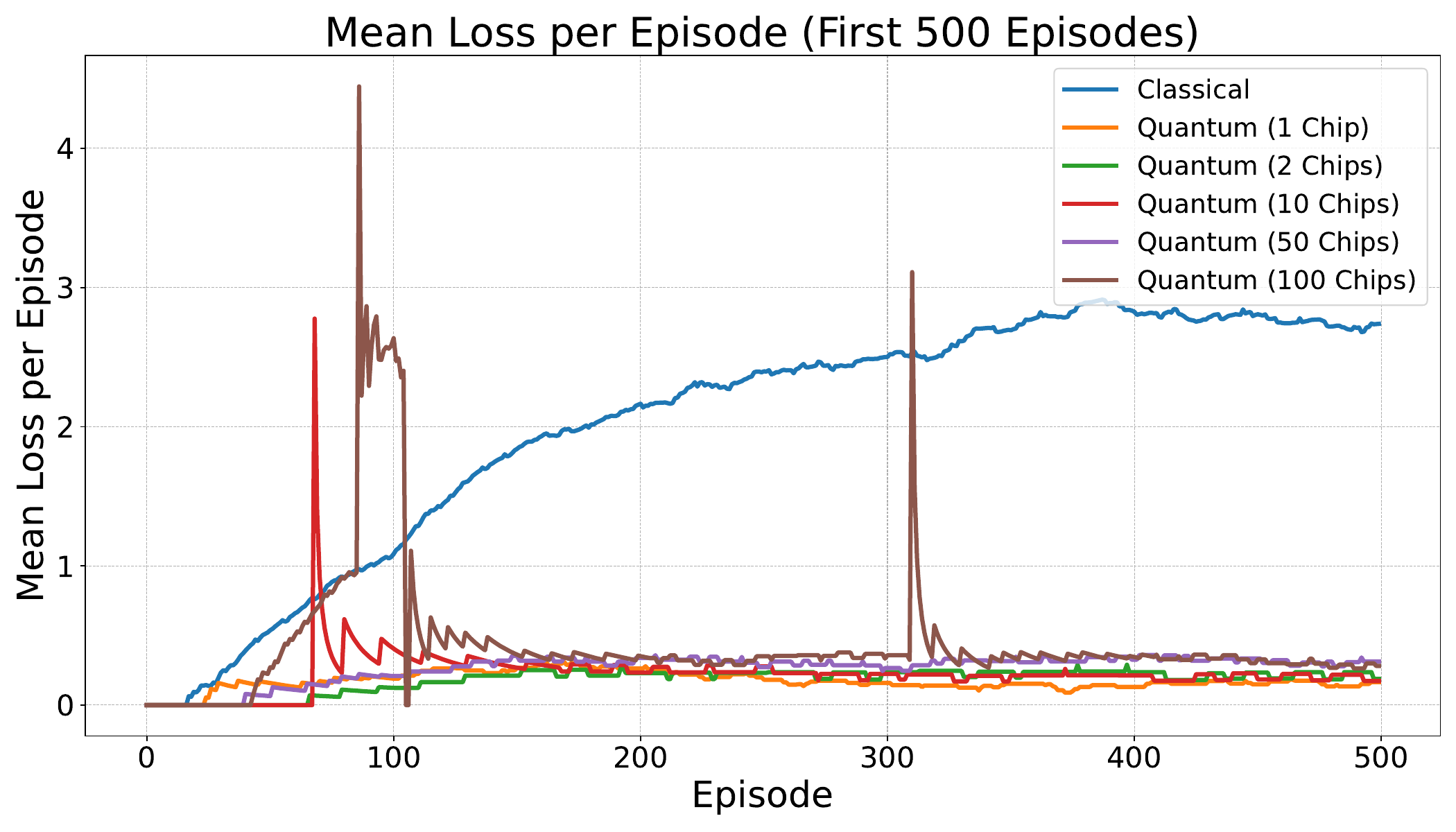}
        \caption{Mean Loss}
        \label{Figure_Loss}
    \end{subfigure}
    \caption{Performance Comparison of Classical, Single-chip, and Multi-chip Ensemble (Q)CNN Agents on Super Mario Bros.}
    \label{Fig_Performance}
\end{figure*}

We present the empirical evaluation of the three RL models: classical baseline, single-chip QCNN, and multi-chip ensemble QCNN models with varying number of quantum chips (2, 10, 50, and 100). The performance was assessed based on the mean reward and mean loss, which are computed as moving average over the last 100 episodes.

Fig. \ref{Figure_Rewards} illustrates the mean reward per episode. The multi-chip Ensemble QCNN models with 50 and 100 chips consistently achieved the highest sustained average rewards, stabilizing around 650-700 after initial rapid learning. The single-chip QCNN model exhibited a sharp initial reward spike exceeding 800 around episode 10-20, but its performance declined significantly thereafter, stabilizing at a lower average reward of approximately 250-300 before a slight recovery. Conversely, the classical model showed a gradual increase in reward, reaching around 400 by episode 500, consistently outperformed by the higher-chip quantum ensembles. Multi-chip ensemble models with fewer chips (2 and 10) also displayed rapid initial learning but showed a decline in sustained reward after episode 150, performing closer to the classical baseline by the end of the observed period.

Fig. \ref{Figure_Loss} depicts the mean loss per episode. All multi-chip ensemble QCNN models, as well as the single-chip QCNN model, demonstrated effective loss minimization after initial transient spikes, converging to very low average loss values, generally below 0.5, from approximately episode 150 onwards. Notably, the classical CNN model exhibited a continuous and significant increase in average loss, reaching nearly 3 by episode 500, indicating persistent difficulty in reducing internal prediction errors. While higher-chip multi-chip ensemble QCNN models (50 and 100 chips) achieved both high rewards and low loss, the single-chip QCNN model, despite its low loss, did not sustain high reward performance, suggesting potential suboptimal policy learning or limitations in exploration despite internal error reduction.

In summary, the multi-chip ensemble QCNN architecture, particularly with 50 and 100 chips, demonstrated superior performance by achieving the highest and most stable average rewards while concurrently minimizing average loss. This suggests an efficient and robust learning process. Although the single-chip QCNN model also achieved low loss, it failed to sustain high rewards. The classical CNN model was consistently outperformed in terms of reward and exhibited a continuously increasing loss, highlighting its limitations for this task.



\section{\textbf{Conclusion}}
We have demonstrated a scalable quantum reinforcement learning framework that enables QRL agents to operate in complex, high-dimensional environments previously inaccessible to quantum approaches. Our multi-chip ensemble QCNN architecture successfully addresses a fundamental limitation in quantum machine learning: the scalability bottleneck caused by limited qubit counts on near-term quantum devices.

Conventional QRL approaches face a critical dilemma: they either require significant classical dimensionality reduction (losing valuable information) or demand impractical quantum resources through techniques like amplitude encoding (requiring exponential circuit depth) \cite{Plesch, Sun2023, Zhang2021}. Our distributed approach resolves this by partitioning input features across multiple small quantum circuits, thereby reducing the dimensional compression burden while keeping individual quantum circuits manageable for NISQ devices \cite{Cerezo2022, Bharti}.

Our experimental results with the Super Mario Bros environment reveal two key findings. First, both quantum approaches (single-chip and multi-chip ensemble) outperformed the classical baseline, achieving higher rewards and lower loss. This suggests potential quantum advantages stemming from the superior expressibility of VQCs~\cite{Du2020} and the capacity of quantum entanglement to explore larger regions of Hilbert space~\cite{Park2024, Zhao}. Second, the multi-chip ensemble QCNN demonstrated superior learning stability and efficiency compared to the single-chip model, with steadier reward improvements and consistently lower loss values throughout training.

Our architecture makes a deliberate trade-off: sacrificing cross-chip entanglement for scalability and NISQ feasibility~\cite{Park2025}. This limits the ability to capture non-local correlations between distant input features within individual quantum circuits. While a single, large-scale quantum processor could theoretically leverage entanglement to discover such correlations more efficiently, our approach remains a practical solution for the NISQ era, trading the potential for a more holistic quantum representation for a significant reduction in quantum hardware requirements.

The multi-chip ensemble architecture offers inherent advantages beyond mere scalability. By distributing computation across smaller, independent quantum circuits, it naturally mitigates barren plateaus through reduced circuit complexity and increased gradient variance \cite{Park2025}. Additionally, this approach provides built-in noise resilience—smaller circuits accumulate less noise \cite{Preskill}, while the classical ensemble averaging reduces output variance from quantum noise \cite{Park2025}. Unlike traditional error mitigation techniques that often trade reduced bias for increased variance \cite{Cai, Temme}, our architectural solution addresses both simultaneously.

Despite these promising results, limitations remain. Our work was conducted in simulation, and performance on actual quantum hardware will face additional challenges from device noise and connectivity constraints. While our approach reduces classical preprocessing requirements, significant classical components remain for initial feature extraction and final decision-making. A deeper theoretical analysis of information flow through this hybrid quantum-classical architecture is still needed. Our classical baseline, while structurally analogous, may not represent the best possible classical methods—comprehensive benchmarking against advanced architectures like deeper residual networks is needed to establish definitive quantum advantage claims.

Future work should focus on hardware implementation to assess real-world viability, exploration of advanced ensemble strategies (hierarchical processing, alternative aggregation functions), investigation of domain-specific quantum circuit designs, application to diverse high-dimensional problems beyond gaming environments, and formal quantification of expressive power and potential quantum advantage in this distributed framework.

This work represents a significant step toward practical quantum reinforcement learning for complex problems, establishing distributed quantum processing as a viable strategy for enhancing near-term quantum capabilities in machine learning applications.

\section*{\textbf{Acknowledgment}}
The authors extend their gratitude to the members of the SNU Connectome Lab for their invaluable support and critical contributions to the development multi-chip ensembles. This work was supported by the National Research Foundation of Korea (NRF) grant funded by the Korea government (MSIT) (No. 2021R1C1C1006503, RS-2023-00266787, RS-2023-00265406, RS-2024-00421268, RS-2024-00342301, RS-2024-00435727, NRF-2021M3E5D2A01022515), by Creative-Pioneering Researchers Program through Seoul National University (No. 200-20240057, 200-20240135), by Semi-Supervised Learning Research Grant by SAMSUNG (No.A0342-20220009), by Identify the network of brain preparation steps for concentration Research Grant by LooxidLabs (No.339-20230001), by Institute of Information \& communications Technology Planning \& Evaluation (IITP) grant funded by the Korea government (MSIT) [NO.RS-2021-II211343, NO.2021-0-01343, Artificial Intelligence Graduate School Program (Seoul National University)] by the MSIT (Ministry of Science, ICT), Korea, under the Global Research Support Program in the Digital Field program (RS-2024-00421268) supervised by the IITP (Institute for Information \& Communications Technology Planning \& Evaluation), by the National Supercomputing Center with supercomputing resources including technical support (KSC-2023-CRE-0568), by the Ministry of Education of the Republic of Korea and the National Research Foundation of Korea (NRF-2021S1A3A2A02090597), by the Korea Health Industry Development Institute (KHIDI), and by the Ministry of Health and Welfare, Republic of Korea (HR22C1605), by Artificial intelligence industrial convergence cluster development project funded by the Ministry of Science and ICT (MSIT, Korea) \& Gwangju Metropolitan City and by KBRI basic research program  through  Korea  Brain  Research  Institute funded by Ministry of Science and ICT (25-BR-05-01).


\bibliographystyle{ieeetr}
\bibliography{references}

\begin{thebibliography}{10}

\bibitem{Chen}
S.~Y.~C. Chen, C.~H.~H. Yang, J.~Qi, P.~Y. Chen, X.~Ma, and H.~S. Goan, ``Variational quantum circuits for deep reinforcement learning,'' {\em IEEE Access}, vol.~8, pp.~141007--141024, 2020.

\bibitem{Jerbi}
S.~Jerbi, C.~Gyurik, S.~Marshall, H.~Briegel, and V.~Dunjko, ``Parametrized quantum policies for reinforcement learning,'' in {\em Advances in Neural Information Processing Systems}, vol.~34, pp.~28362--28375, 2021.

\bibitem{Skolik}
A.~Skolik, S.~Jerbi, and V.~Dunjko, ``Quantum agents in the gym: a variational quantum algorithm for deep q-learning,'' {\em Quantum}, vol.~6, p.~720, 2022.

\bibitem{Park2024}
J.~Park, J.~Cha, S.~Y.-C. Chen, S.~Yoo, and H.-H. Tseng, ``Over the quantum rainbow: Explaining hybrid quantum reinforcement learning,'' in {\em 2024 IEEE International Conference on Quantum Computing and Engineering (QCE)}, vol.~01, pp.~1583--1594, 2024.

\bibitem{Lockwood2020}
O.~Lockwood and M.~Si, ``Reinforcement learning with quantum variational circuit,'' {\em Proceedings of the AAAI Conference on Artificial Intelligence and Interactive Digital Entertainment}, vol.~16, pp.~245--251, Oct. 2020.

\bibitem{Bharti}
K.~Bharti, A.~Cervera-Lierta, T.~H. Kyaw, T.~Haug, S.~Alperin-Lea, A.~Anand, {\em et~al.}, ``Noisy intermediate-scale quantum algorithms,'' {\em Rev. Mod. Phys.}, vol.~94, p.~015004, Feb 2022.

\bibitem{Plesch}
M.~Plesch and i.~c.~v. Brukner, ``Quantum-state preparation with universal gate decompositions,'' {\em Phys. Rev. A}, vol.~83, p.~032302, Mar 2011.

\bibitem{Sun2023}
X.~Sun, G.~Tian, S.~Yang, P.~Yuan, and S.~Zhang, ``Asymptotically optimal circuit depth for quantum state preparation and general unitary synthesis,'' {\em IEEE Transactions on Computer-Aided Design of Integrated Circuits and Systems}, vol.~42, no.~10, pp.~3301--3314, 2023.

\bibitem{Zhang2021}
X.-M. Zhang, M.-H. Yung, and X.~Yuan, ``Low-depth quantum state preparation,'' {\em Phys. Rev. Res.}, vol.~3, p.~043200, Dec 2021.

\bibitem{DePalma}
G.~De~Palma, M.~Marvian, C.~Rouz\'e, and D.~S. Fran\ifmmode~\mbox{\c{c}}\else \c{c}\fi{}a, ``Limitations of variational quantum algorithms: A quantum optimal transport approach,'' {\em PRX Quantum}, vol.~4, p.~010309, Jan 2023.

\bibitem{Lockwood2021}
O.~Lockwood and M.~Si, ``Playing {A}tari with hybrid quantum-classical reinforcement learning,'' in {\em NeurIPS 2020 Workshop on Pre-registration in Machine Learning}, pp.~285--301, PMLR.

\bibitem{McClean}
J.~R. McClean, S.~Boixo, V.~N. Smelyanskiy, R.~Babbush, and H.~Neven, ``Barren plateaus in quantum neural network training landscapes,'' {\em Nature Communications}, vol.~9, no.~1, p.~4812, 2018.

\bibitem{Larocca}
M.~Larocca, S.~Thanasilp, S.~Wang, K.~Sharma, J.~Biamonte, P.~J. Coles, {\em et~al.}, ``Barren plateaus in variational quantum computing,'' {\em Nature Reviews Physics}, vol.~7, no.~4, pp.~174--189, 2025.

\bibitem{Kimble}
H.~J. Kimble, ``The quantum internet,'' {\em Nature}, vol.~453, no.~7198, pp.~1023--1030, 2008.

\bibitem{Peng}
T.~Peng, A.~W. Harrow, M.~Ozols, and X.~Wu, ``Simulating large quantum circuits on a small quantum computer,'' {\em Phys. Rev. Lett.}, vol.~125, p.~150504, Oct 2020.

\bibitem{Barral}
D.~Barral, F.~J. Cardama, G.~Díaz-Camacho, D.~Faílde, I.~F. Llovo, M.~Mussa-Juane, {\em et~al.}, ``Review of distributed quantum computing: From single qpu to high performance quantum computing,'' {\em Computer Science Review}, vol.~57, p.~100747, 2025.

\bibitem{Kawase}
Y.~Kawase, ``Distributed quantum neural networks via partitioned features encoding,'' {\em Quantum Machine Intelligence}, vol.~6, no.~1, p.~15, 2024.

\bibitem{Wu}
J.~Wu, T.~Hu, and Q.~Li, ``Distributed quantum machine learning: Federated and model-parallel approaches,'' {\em IEEE Internet Computing}, vol.~28, no.~2, pp.~65--72, 2024.

\bibitem{Pira}
L.~Pira and C.~Ferrie, ``An invitation to distributed quantum neural networks,'' {\em Quantum Machine Intelligence}, vol.~5, no.~2, p.~23, 2023.

\bibitem{QUDIO}
Y.~Du, Y.~Qian, X.~Wu, and D.~Tao, ``A distributed learning scheme for variational quantum algorithms,'' {\em IEEE Transactions on Quantum Engineering}, vol.~3, pp.~1--16, 2022.

\bibitem{Park2025}
J.~J. Park, J.~Cha, S.~Y.-C. Chen, H.-H. Tseng, and S.~Yoo, ``Addressing the current challenges of quantum machine learning through multi-chip ensembles,'' 2025.

\bibitem{SuperMarioBros}
C.~Kauten, ``{S}uper {M}ario {B}ros for {O}pen{AI} {G}ym.'' GitHub, 2018.

\bibitem{Bellman}
R.~E. Bellman, {\em Dynamic Programming}.
\newblock Princeton, NJ: Princeton University Press, 1957.

\bibitem{Mnih}
V.~Mnih, K.~Kavukcuoglu, D.~Silver, A.~A. Rusu, J.~Veness, M.~G. Bellemare, {\em et~al.}, ``Human-level control through deep reinforcement learning,'' {\em Nature}, vol.~518, no.~7540, pp.~529--533, 2015.

\bibitem{vanHasselt}
H.~van Hasselt, A.~Guez, and D.~Silver, ``Deep reinforcement learning with double q-learning,'' {\em Proceedings of the AAAI Conference on Artificial Intelligence}, vol.~30, no.~1, 2016.

\bibitem{vanHasseltHado}
H.~van Hasselt, ``Double {Q}-learning,'' {\em Advances in Neural Information Processing Systems 23 (NIPS 2010)}, vol.~23, p.~2613–2621, 2010.

\bibitem{Benedetti}
M.~Benedetti, E.~Lloyd, S.~Sack, and M.~Fiorentini, ``Parameterized quantum circuits as machine learning models,'' {\em Quantum Science and Technology}, vol.~4, no.~4, p.~043001, 2019.

\bibitem{Schuld2019}
M.~Schuld, V.~Bergholm, C.~Gogolin, J.~Izaac, and N.~Killoran, ``Evaluating analytic gradients on quantum hardware,'' {\em Phys. Rev. A}, vol.~99, p.~032331, Mar 2019.

\bibitem{Cong}
I.~Cong, S.~Choi, and M.~D. Lukin, ``Quantum convolutional neural networks,'' {\em Nature Physics}, vol.~15, no.~12, pp.~1273--1278, 2019.

\bibitem{Pesah}
A.~Pesah, M.~Cerezo, S.~Wang, T.~Volkoff, A.~T. Sornborger, and P.~J. Coles, ``Absence of barren plateaus in quantum convolutional neural networks,'' {\em Phys. Rev. X}, vol.~11, p.~041011, Oct 2021.

\bibitem{Kerenidis}
I.~Kerenidis, J.~Landman, and A.~Prakash, ``Quantum algorithms for deep convolutional neural networks,'' in {\em International Conference on Learning Representations}, 2020.

\bibitem{bergholm2022pennylane}
V.~Bergholm, J.~Izaac, M.~Schuld, C.~Gogolin, S.~Ahmed, V.~Ajith, {\em et~al.}, ``Pennylane: Automatic differentiation of hybrid quantum-classical computations,'' 2022.

\bibitem{Cerezo2022}
M.~Cerezo, G.~Verdon, H.-Y. Huang, L.~Cincio, and P.~J. Coles, ``Challenges and opportunities in quantum machine learning,'' {\em Nature Computational Science}, vol.~2, no.~9, pp.~567--576, 2022.

\bibitem{Du2020}
Y.~Du, M.-H. Hsieh, T.~Liu, and D.~Tao, ``Expressive power of parametrized quantum circuits,'' {\em Phys. Rev. Res.}, vol.~2, p.~033125, Jul 2020.

\bibitem{Zhao}
H.~Zhao and D.-L. Deng, ``Entanglement-induced provable and robust quantum learning advantages,'' 2024.

\bibitem{Preskill}
J.~Preskill, ``Quantum {C}omputing in the {NISQ} era and beyond,'' {\em {Quantum}}, vol.~2, p.~79, Aug. 2018.

\bibitem{Cai}
Z.~Cai, R.~Babbush, S.~C. Benjamin, S.~Endo, W.~J. Huggins, Y.~Li, {\em et~al.}, ``Quantum error mitigation,'' {\em Rev. Mod. Phys.}, vol.~95, p.~045005, Dec 2023.

\bibitem{Temme}
K.~Temme, S.~Bravyi, and J.~M. Gambetta, ``Error mitigation for short-depth quantum circuits,'' {\em Phys. Rev. Lett.}, vol.~119, p.~180509, Nov 2017.

\end{thebibliography}

\end{document}